\documentstyle[aps]{revtex}
\draft
\begin{document}
\newcommand{\fiv}{\frac{1}{T_{1}}\phi^{m}_{T_{1}}(v)}
\newcommand{\Fiv}{\frac{1}{T_{2}}\phi^{m}_{T_{2}}(v)}
\newcommand{\fiV}{\frac{1}{T_{1}}\phi^{m}_{T_{1}}(V)}
\newcommand{\FiV}{\frac{1}{T_{2}}\phi^{m}_{T_{2}}(V)}
\newcommand{\ilV}{\int_{-\infty}^{V}dv}
\newcommand{\igV}{\int_{V}^{\infty}dv}
\newcommand{\be}{\begin{equation}}
\newcommand{\ee}{\end{equation}}
\newcommand{\PH}{\Phi_{\epsilon}}
\newcommand{\FE}{F_{\epsilon}(x)}

\title{Stationary Motion of the Adiabatic Piston}

\author{Ch.\ Gruber}
\address{Institut de Physique Th\'{e}orique, \'Ecole Polytechnique
F\'ed\'erale de Lausanne, CH-1015 Lausanne, Switzerland}
\author{J.\ Piasecki}
\address{Institute of Theoretical Physics, University of Warsaw,
Ho\.{z}a 69, PL-00 681 Warsaw, Poland}

\date{\today}

\maketitle

\begin{abstract}

We consider a one-dimensional system consisting of two infinite ideal fluids, with equal pressures but different temperatures $T_{1}$ and $T_{2}$, separated by an adiabatic movable piston whose mass $M$ is much larger than the mass $m$ of the fluid particules. This is the infinite version of the controversial adiabatic piston problem. The stationary non-equilibrium solution of the Boltzmann equation for the velocity distribution of the piston is expressed in powers of the small parameter $\epsilon=\sqrt{m/M}$, and explicitly given up to order $\epsilon^{2}$. In particular it implies that although the pressures are equal on both sides of the piston, the temperature difference induces a non-zero average velocity of the piston in the direction of the higher temperature region. It thus shows that the asymmetry of the fluctuations induces a macroscopic motion despite the absence of any macroscopic force. This same conclusion was previously obtained for the non-physical situation where $M=m$. 
\end{abstract}


\section{Introduction}

The {\it adiabatic piston problem} has been the subject of continuous controversy for many years (see [1] and references given therein). The problem is the following. The system is a finite cylinder containing two gases separated by an adiabatic movable piston. Initially the piston is fixed by a brake and the two gases are in equilibrium. At a certain time the brake is released and the question is to find the final equilibrium state. Formulated differently, it is one of the ``Problems in statistical mechanics that I would like to see solved'', mentionned by E. Lieb in his talk at Statphys 20 [2]. The controversy is related to the fact that it is a well known example, discussed in many text books (e.g. [3]), for which the principle of entropy maximum, i.e the $2^{\rm nd}$ law of thermostatics, is not sufficient to find the equilibrium position. Indeed the equilibrium position depends in an essential manner on the viscous force and one has to solve explicitly the time evolution to find the final position (just like in mechanics when solid friction is present). In other words it is a problem which can be solved within the framework of thermodynamics (of irreversible processes), but not within the framework of thermostatics [1].

In this context the question has often been raised whether the adiabatic condition is an ideal constraint, which is essential to discuss the foundation of thermodynamics [4], but which can never be realized in a real physical situation (just as the condition of a frictionless piston). The question has even been extended to whether the $2^{\rm nd}$ law does not apply in certain situation.

As we know, heat conduction through a wall is associated with the internal motion of the molecules of the wall. If the wall is a rigid ideal solid which is fixed, the atoms of the gas will make purely elastic collisions on the wall and it is possible to maintain at different temperatures two gases separated by such a wall. The wall is thus adiabatic. The problem is then to analyse what happens when this same ideal wall is no longer fixed, but is allowed to move. The question is whether the stochastic motion of the wall, induced by the fluctuations in the force exerted by the particules of the gas, will lead to an equilibrium state in which the temperatures of both gases are equals. In other words whether an adiabatic fixed wall becomes diathermic when it is allowed to move.

Recently this problem was investigated using the following model. The adiabatic piston was considered as a heavy, structureless, particule of mass $M$ submitted on both sides to purely elastic collisons with the atoms of two gases in equilibrium with temperatures $T_{1}$ and $T_{2}$ respectively [5]. It was assumed that both gases are ideal and made of identical point particules of mass $m$. Considering first that $M \gg m$, a linear Boltzmann equation was derived for the velocity distribution of the piston. The rigorous stationary solution was then found for the special case where one takes $M=m$ in the Boltzmann equation and the two gases extend to infinity, respectively to the left and to the right of the piston. The main conclusion was that the fluctuations conspire with the spatial anisotropy ($T_{1} \neq T_{2}$) to induce a macroscopic motion of the piston, with finite average velocity in the direction of the higher temperature region, although no macroscopic force is present (the pressures on both sides of the piston are equal). The same model with $M=m$ was then considered in [6] for a system with finite length; it was shown that the piston evolves toward an equilibrium position corresponding to a uniform density of particules throughout the system.

In the following we study the more physical situation where $M \gg m$. In this case the Boltzmann equation describes asymptotically exactly the dynamic of the piston. This is because the massive piston has a vanishing probability to interact back with the perturbation it causes in the states of the surrounding fluids (no recollisions in the sense of the kinetic
theory). These perturbations go away to infinity, and the piston at consecutive collisions always "sees" on both sides the unperturbed initial equilibrium states. This same model was alredy considered by J.L. Lebowitz [7]. 

In the special case of equal temperatures on both sides of the piston, i.e $T_{1}=T_{2}$, the problem reduces to the classical Brownian motion of the ``Rayleigh piston'', which has been extensively studied in the literature [8-11], using for example the Kramers-Moyal expansion in powers of $1/\sqrt{M}$ [12]. The general problem of a heavy particule undergoing elastic collisons with an infinite ideal gas of light atoms in equilibrium has been rigorously treated by Holley in one dimension [13], and by D\"urr, Goldstein, Lebowitz [13], followed by Goldstein, Guetti [14] in arbitrary dimension. In the limit where $m\rightarrow 0$ and the density of the gas in proportional to $m^{-1/2}$, these authors established the convergence of the mechanical process describing the velocity of the heavy molecule to the Ornstein-Uehlenbeck process. Therefore when the temperature of the bath is uniform the heavy particule will finally thermalized, acquiring the temperature of the surrounding thermostat. The characteristic velocities of the piston and of the gas particles would be $\sqrt{k_{B}T/M}$ and $\sqrt{k_{B}T/m}$, respectively. So, as a natural small parameter for our investigation with $T_{1} \neq T_{2}$ we choose
\be
\epsilon \; =\; \sqrt{\frac{m}{M}} \label{6}
\ee
The mechanical model is presented in section II. In section III we develop the perturbative analysis of the stationary solution of the Boltzmann equation for $\epsilon \to 0$. Section IV contains the evaluation of the average velocity of the piston and concluding comments.

\section{Mechanical model}

We consider a gas composed of identical point particles of mass $m$ moving in
$R^{1}$. The motion between collisions is free. When two particles collide
they instantaneously exchange their velocities, the collision being
perfectly elastic.

Up to the initial moment $t=t_{0}$ a particle of mass $M$, fixed at some 
point, isolates the volume of the gas to its left  (subsystem
$1$) from that to its right (subsystem $2$). Collisions with the fixed mass
$M$ are assumed to be elastic (specular reflections). 
The two infinite volumes of
the gas on both sides of the mass $M$ are at thermal equilibrium characterized
by the number densities 
$n_{1},\; n_{2}$ and temperatures $T_{1},\; T_{2}$, respectively. The
equilibrium pressures are given by the perfect gas equation of state
\be
p_{i}\; = \; n_{i}k_{B}T_{i}, \;\;\; i=1,2, \label{1}
\ee
where $k_{B}$ is the Boltzmann constant.

At the moment $t=t_{0}$ the constraint fixing the mass $M$ is removed,
and the released mass starts to move freely suffering elastic collisions with
the gas particles.
If before the collision the masses $M$ and $m$ have velocities $V$ and
$v$, respectively, their velocities after the collision are given by
\begin{eqnarray}
V' & = & V - \frac{2m}{m+M} (V-v) \label{2}\\
v' & = & v + \frac{2M}{m+M} (V-v)\nonumber
\end{eqnarray}

Our object here is to determine the asymptotic stationary velocity 
distribution of the mass $M$ in the case where $M >> m$, assuming the
initial pressures (\ref{1}) to be equal
\be
p_{1}\; =\; p_{2}\; =\; p \label{3}
\ee 
From the physical point of view we thus consider an infinite 
volume version 
of the {\it adiabatic piston problem} discussed in the introduction and which is essentially a one-dimensional problem. The massive 
structureless particle plays here the role of the movable adiabatic wall (or 
piston), and the fundamental question concerns the nature of the asymptotic 
stationary state. The assumed equality of the initial pressures
(\ref{3}) represents the compensation of the momentum fluxes arriving at the 
fixed piston
\be
p_{1}\; =\; \int_{0}^{\infty}dv\, v\, 2mv \phi^{m}_{T_{1}}(v)
\; =\; \int_{-\infty}^{0}dv\, v\,(-2mv) \phi^{m}_{T_{2}}(v)\; =\; p_{2}
\label{4}
\ee   
where $\phi^{m}_{T}$ denotes the Maxwell distribution
\be
\phi^{m}_{T}(v) = \sqrt{\frac{m}{2\pi k_{B}T}}{\rm exp}\left(
                   -\frac{mv^{2}}{k_{B}T}\right) \label{5}
\ee
The average motion of the released piston will thus reflect the effects of
fluctuations in the momentum transfers at binary encounters.

\section{Perturbative analysis of the stationary state}

As discussed in [5], the velocity distribution function
$\Phi_{\epsilon}(V)$ of the piston in the stationary state with equal pressures
on both sides is given by the solution of the linear Boltzmann equation. 
With the use of the Dirac $\delta$-distribution we can write it in the form
\begin{eqnarray}
\label{7}
 \int dV' \int &d&v' (v'-V')\theta (v'-V')\frac{1}{T_{1}}\phi^{m}_{T_{1}}(v')
\Phi_{\epsilon}(V')\delta \left(V'+\frac{2m}{m+M}(v'-V')-V \right) \nonumber\\
- \int  &d&v (v-V)\theta (v-V)\fiv \Phi_{\epsilon}(V) \\
 +\int dV' \int &d&v' (V'-v')\theta (V'-v')\frac{1}{T_{2}}\phi^{m}_{T_{2}}(v')
\Phi_{\epsilon}(V')\delta\left(V'+\frac{2m}{m+M}(v'-V')-V \right) \nonumber\\
-\int &d&v (V-v)\theta (V-v)\Fiv \Phi_{\epsilon}(V)\;\; = \;\; 0 \nonumber
\end{eqnarray}

Equation (\ref{7}) describes the situation where the temperatures $T_{1}$ and
$T_{2}$ characterize the initial states to the left and to the right of
the piston, respectively.

With the small parameter $\epsilon = \sqrt{m/M}$ introduced in section I we have from
equations (\ref{2}) 
\be
v'-V' = \frac{1+\epsilon^{2}}{1-\epsilon^{2}}(v'-V) \label{8}
\ee
So, the integration over $V'$ in (\ref{7}) yields the equation

\begin{eqnarray}
\label{9}
 \left( \frac{1+\epsilon^{2}}{1-\epsilon^{2}}\right)^{2}
&&\igV (v-V)\fiv \PH
\left[V-\frac{2\epsilon^{2}}{1-\epsilon^{2}}(v-V)\right]\nonumber\\
- &&\igV (v-V)\fiv \PH (V) \\ 
 + \left( \frac{1+\epsilon^{2}}{1-\epsilon^{2}}\right)^{2}&&\ilV (V-v)\Fiv
\PH \left[V+\frac{2\epsilon^{2}}{1-\epsilon^{2}}(V-v)\right] \nonumber \\
 -&&\ilV (V-v)\Fiv \PH (V) \;\; =\;\; 0 \nonumber
 \end{eqnarray}

which can be conveniently rewritten in the form
\begin{eqnarray}
\label{10}
&&\left[\left( \frac{1+\epsilon^{2}}{1-\epsilon^{2}}\right)^{2}-1\right]
\left\{ \igV (v-V)\fiv + \ilV (V-v) \Fiv \right\}\PH (V)\nonumber \\
 + &&\left( \frac{1+\epsilon^{2}}{1-\epsilon^{2}}\right)^{2} \igV (v-V)\fiv
 \left\{\PH[V-\frac{2\epsilon^{2}}{1-\epsilon^{2}}(v-V)] -\PH (V) \right\}\nonumber\\
 + &&\left( \frac{1+\epsilon^{2}}{1-\epsilon^{2}}\right)^{2}
\ilV (V-v)\Fiv \left\{\PH[V-\frac{2\epsilon^{2}}{1-\epsilon^{2}}(v-V)] -\PH (V)\right\}
\; =\; 0
\end{eqnarray}
Using the expansion of $\PH [V-2\epsilon^{2}(v-V)/(1-\epsilon^{2})]$ in
powers 
\begin{eqnarray}
\frac{2\epsilon^{2}(v-V)}{(1-\epsilon^{2})}
\end{eqnarray}
 we obtain

 \begin{eqnarray}
 \label{11}
 0 &=&  2 \left\{\igV (v-V)\fiv + \ilV (V-v) \Fiv \right\}\PH (V) \\
 &+& (1+\epsilon^{2})^{2} \sum_{n=1}^{\infty}\frac{(-1)^{n}}{n!}
\frac{(2\epsilon^{2})^{n-1}}{(1-\epsilon^{2})^{n}}\igV  (v-V)^{n+1}
\fiv \left(\frac{d}{dV}\right)^{n}\PH (V) \nonumber \\
 &+&(1+\epsilon^{2})^{2} \sum_{n=1}^{\infty}\frac{1}{n!}
\frac{(2\epsilon^{2})^{n-1}}{(1-\epsilon^{2})^{n}}\ilV  (V-v)^{n+1}
\Fiv \left(\frac{d}{dV}\right)^{n}\PH (V)\nonumber
\end{eqnarray}
The series in (\ref{11}) corresponds to the $\epsilon$-expansion of the
collision law (\ref{2}).

Introducing the functions $I_{n}(V)$ defined by
\be
I_{n}(V)=\frac{1}{n!}\left\{ \ilV (V-v)^{n}\Fiv -
(-1)^{n}\igV (v-V)^{n}\fiv  \right\} \label{12}
\ee
we get from (\ref{11}) the following equation for $\PH (V)$
\be
0\;\; = \;\; 2I_{1}(V)\PH (V)+(1+\epsilon^{2})^{2} \sum_{n=1}^{\infty}
\frac{(2\epsilon^{2})^{n-1}}{(1-\epsilon^{2})^{n}}(n+1)I_{n+1}(V) 
\left(\frac{d}{dV}\right)^{n}\PH (V)  \label{13}
\ee
Let us first establish some useful properties of the functions $I_{n}(V)$.
From the definition (\ref{12}), we have
\be
\label{more}
\frac{d}{dV}I_{n}(V)\; = \; I_{n-1}(V),\;\; n=0,1,2,... \label{14}
\ee
We define $I_{-1} $ by
\begin{eqnarray}
I_{-1}(V) & = & \frac{d}{dV}I_{0}(V) = \fiV + \FiV \label{15} \\
{} & = &  \sqrt{\frac{m}{2\pi k_{B}}}\left[ \frac{1}{T_{1}^{3/2}}
{\rm exp}\left(-\frac{mV^{2}}{k_{B}T_{1}}\right) 
+ \frac{1}{T_{2}^{3/2}} {\rm exp}\left( -\frac{mV^{2}}{k_{B}T_{2}}\right)
\right] \nonumber
\end{eqnarray}

It will be convenient to introduce the expansion of $I_{n}(V)$ in powers of $V$
\begin{equation}
\label{16}
I_n(V)=\sum_{q=1}^{\infty}\frac{1}{q!}I_{n,q}\,{V^{q}}
\end{equation}
Using equation (\ref{more}) yields
\begin{eqnarray}
\label{17}
I_{n,q}\; &=& \; I_{n-1,q-1} \Longrightarrow \\
\label{18}
I_{n,q}\; &=& \; I_{n-q,0}, \;\;\forall q\leq n \\
\label{19}
I_{n,q}\; &=& \; I_{-1,q-n-1} ,\;\; \forall q\geq n+1
\end{eqnarray}
From (\ref{15}) and (\ref{18}) we obtain for $ q\geq n+1$
\be
I_{n,n+2l} = 0 \label{20}
\ee
and
\be
I_{n,n+2l+1} = \frac{(-1)^{l}}{\sqrt{\pi}}\left(\frac{m}{2k_{B}}
\right)^{l+1/2}\left( \frac{1}{T_{2}^{l+3/2}}+\frac{1}{T_{1}^{l+3/2}}
\right) \label{21}
\ee
On the other hand, for $q\leq n$, the relation (\ref{18}) yields $I_{n,q} = I_{n-q}(V=0)$ and
\be
I_{n,q} = \frac{1}{2^{n-q+1}\Gamma (1+(n-q)/2)}\left(\frac{2k}{m} \right)^{(n-q)/2}
[ T_{2}^{(n-q-2)/2} - (-1)^{n-q}T_{1}^{(n-q-2)/2}] \label{22}
\ee
with
\begin{eqnarray}
\Gamma (1+l) & = & l! \label{23} \\
\Gamma (1+\frac{2k+1}{2}) & = & \frac{\sqrt{\pi}}{2^{k+1}}1\cdot 3\cdot
5\, ... \,\cdot (2k+1) \nonumber
\end{eqnarray}
Using the relation (\ref{14}), one can rewrite equation (\ref{13})  
in the form
\be
\frac{d}{dV}\left\{ \sum_{n=0}^{\infty}\left(\frac{2\epsilon^{2}}{1-\epsilon^{2}}\right)^{n}
(2+n+n\epsilon^{2})I_{2+n}(V)\left(\frac{d}{dV}\right)^{n}\PH(V)\right\}
= 0 \label{24}
\ee

Integrating (\ref{24}) over the interval $[V,\infty )$, and assuming that
the integrand vanishes in the limit $V\to \infty$, we obtain
\be
 \sum_{n=0}^{\infty}\left(\frac{2\epsilon^{2}}{1-\epsilon^{2}}\right)^{n}
(2+n+n\epsilon^{2})I_{2+n}(V)\left(\frac{d}{dV}\right)^{n}\PH(V)= 0 
\label{240}
\ee

For $T_{1}=T_{2}=T$, the solution of (\ref{10}), or (\ref{240}), is the
Maxwell distribution
\be
\PH (V)|_{T_{1}=T_{2}=T} = \phi^{M}_{T}(V)=\sqrt{\frac{M}{2\pi k_{B}T}}
{\rm exp}\left(-\frac{MV^{2}}{2k_{B}T}\right) \label{25}
\ee
This suggests to introduce the dimensionless variable
\be
x = \sqrt{\frac{M}{2k_{B}\sqrt{T_{1}T_{2}}}}V  \label{26}
\ee
and the normalized distribution $F_{\epsilon}(x)$ defined by
\be
F_{\epsilon}(x)=\sqrt{\frac{2k_{B}\sqrt{T_{1}T_{2}}}{M}}\PH \left(
\sqrt{\frac{2k_{B}\sqrt{T_{1}T_{2}}}{M}}x\right) \label{27}
\ee
We thus have
\be
\left(\frac{d}{dV}\right)^{n}\PH (V) = 
\left(\frac{M}{2k_{B}\sqrt{T_{1}T_{2}}}\right)^{n/2}
\left(\frac{d}{dx}\right)^{n}F_{\epsilon}(x) \label{28}
\ee
Using the expansion (\ref{16}) for $I_{2+n}(V)$ and the change of
variables (\ref{26}) we find that the distribution $F_{\epsilon}(x)$
is the solution of the equation
\be
\sum_{n=0}^{\infty}\sum_{q=0}^{\infty}\left(\frac{2\epsilon^{2}}
{1-\epsilon^{2}}\right)^{n}(2+n+n\epsilon^{2})
\frac{I_{2+n,q}}{q!}  \left[  \frac{2k_{B}}{M}\sqrt{T_{1}T_{2}}
\right]^{\frac{q-n}{2}}x^{q}\left(\frac{d}{dx}\right)^{n}F_{\epsilon}(x)=0
\label{29}
\ee
If $(n-q)\geq -2$, equation (\ref{22}) implies
\begin{eqnarray}
\left[ \frac{2k_{B}}{M}\sqrt{T_{1}T_{2}}\right]^{(q-n)/2}I_{2+n,q} & = & {}
\label{30}\\
 {} & = & \left[ \left( \frac{T_{2}}{T_{1}}\right)^{(n-q)/4} - (-1)^{n-q}
\left( \frac{T_{1}}{T_{2}}\right)^{(n-q)/4}\right]\epsilon^{q-n-2}
\frac{2k_{B}}{M}\frac{2^{q-n-3}}{\Gamma (1+\frac{2+n-q}{2})} \nonumber
\end{eqnarray}
while for $q=n+3+2l,\;\; l=0,1,2,...$, we have from (\ref{21}) 
\begin{eqnarray} 
\hspace{-1.3cm}\left[ \frac{2k_{B}}{M}\sqrt{T_{1}T_{2}}\right]^{(q-n)/2}I_{2+n,q} & = & {}
\label{31}\\
& {} = & \left[ \left( \frac{T_{2}}{T_{1}}\right)^{(n-q)/4}+
\left( \frac{T_{1}}{T_{2}}\right)^{(n-q)/4}\right]\epsilon^{q-n-2}
\frac{2k_{B}}{M}\frac{(-1)^{(q-n-3)/2}}{\sqrt{\pi}} \nonumber
\end{eqnarray}
We thus obtain the following equation for $\FE$
\be
\sum_{n=0}^{\infty}\sum_{q=0}^{\infty}\frac{\epsilon^{n+q}}
{(1-\epsilon^{2})^{n}}(2+n+n\epsilon^{2})2^{n}
J_{n-q}\frac{x^{q}}{q!}\left(\frac{d}{dx}\right)^{n}F_{\epsilon}(x)=0
\label{32}
\ee
where in terms of the parameter 
\be
 \tau = \left(\frac{T_{2}}{T_{1}}\right)^{1/4} \label{33}
\ee
we have
\begin{eqnarray}
J_{r} & = & \frac{[\tau^{r}-(-\tau )^{-r}]}{2^{r+3}
\Gamma (1+(r+2)/2)},\;\; {\rm for} \;\; r\geq -2 \label{34}\\
J_{r} & = & [\tau^{r}+\tau^{-r}]\frac{(-1)^{l}}{\sqrt{\pi}},\;\; {\rm for}
\;\; r=-3-2l,\; l=0,1,2,... \nonumber \\
J_{r} & = & 0, \;\; {\rm for}\;\; r=-4-2l,\;\; l=0,1,2,...
\nonumber
\end{eqnarray}
Let us remark that since $J_{0}=0$, the left hand side of (\ref{32}) vanishes
for $\epsilon =0$, which is consistent with the assumption made when
deriving equation (\ref{240}). Therefore, we can divide (\ref{32})
by $\epsilon$. Finally, assuming that $\FE$ has an asymptotic expansion
\be
\FE  = \sum_{l=0}^{\infty}\epsilon^{l}F_{l}(x)  \label{35}
\ee
in a series of integrable functions $F_{l}(x)$, we are led to solve at 
successive orders in $\epsilon$ the equation
\be
\sum_{n=0}^{\infty}\sum_{q=0}^{\infty}\sum_{m=0}^{\infty}\sum_{l=0}^{\infty}
\epsilon^{n+q+2m+l-1}\frac{(n-1+m)!}{(n-1)!m!}(2+n+n\epsilon^{2})
2^{n}J_{n-q}\frac{x^{q}}{q!}\left(\frac{d}{dx}\right)^{n}F_{l}(x) = 0
\label{36}
\ee
The normalization in (\ref{35}) is supposed to be entirely contained in
the zero order term
\be
\int dx \FE = \int dx F_{0}(x)=1 \label{370}
\ee
We will thus require that 
\be
\int dx F_{r}(x)=0, \;\;\; {\rm for}\;\;\; r=1,2,...  \label{380}
\ee

To obtain the solution for (\ref{36}) at successive orders $r=0,1,2$, 
we give below the explicit form of $J_{r}$
\begin{eqnarray}
J_{1} & = & (\tau + \tau^{-1})/12\sqrt{\pi},\;\;\;\;\;  
J_{-1}=(\tau + \tau^{-1})/2\sqrt{\pi} \label{37} \\
J_{2} & = & (\tau^{2} + \tau^{-2})/26, \;\;\;\;\;\;\;\;
J_{-2}=(\tau^{-2} - \tau^{2})/2 \nonumber \\
J_{3} & =( & \tau^{3} + \tau^{-3})/120\sqrt{\pi}, \;\;  
J_{-3}=(\tau^{3} + \tau^{-3})/\sqrt{\pi} \nonumber
\end{eqnarray}
The terms in (\ref{36}) which contribute to the order $\epsilon^{r}$ 
correspond to the values of the summation variables $n,q,m,l$ satisfying the
relation
\be
n+q+2m+l-1=r  \label{38}
\ee
The last of equations (\ref{34}) shows that some sets satisfying (\ref{38})
yield a vanishing contribution. 

Using equation (\ref{34}) we can now construct the expansion (\ref{35}) in a
systematic way.

\underline{Order $\epsilon^{0}$}\\
When $r=0$, we find
\be
\left[ 2J_{-1}x + 6J_{1}\frac{d}{dx}\right] F_{0}(x) =  
2J_{-1} \left[ x + \frac{1}{2}\frac{d}{dx}\right]F_{0}(x) = 0 \label{39}
\ee
Hence the normalized solution reads
\be
F_{0}(x) = {\rm exp}(-x^{2})/\sqrt{\pi} \label{40}
\ee
We thus have at the order zero (see equation (\ref{27})) a gaussian 
probability density
\be
\Phi_{0}(V)=\sqrt{\frac{M}{2\pi k_{B}\sqrt{T_{1}T_{2}}}}
{\rm exp}\left(-\frac{MV^{2}}{2k_{B}\sqrt{T_{1}T_{2}}}\right) \label{400}
\ee

\underline{Order $\epsilon^{1}$}\\

When $r=1$, equation (\ref{36}) yields a relation between $F_{1}(x)$ and
$F_{0}(x)$ of the form
\be
\left[ 2J_{-1}x + 6J_{1}\frac{d}{dx}\right] F_{1}(x)
+\left[ J_{-2}x^{2}+ 16J_{2} \left(\frac{d}{dx} \right)^{2}\right]F_{0}(x)
=0 \label{41}
\ee
Using (\ref{34}) we can rewrite (\ref{41}) as
\be
\left( x+\frac{1}{2}\frac{d}{dx}\right)F_{1}=
\sqrt{\pi}(\tau - \tau^{-1})\left[ \frac{x^{2}}{2}-
\frac{1}{4}\left(\frac{d}{dx}\right)^{2}\right]F_{0}(x) \label{42}
\ee
The condition
\be
 \int dx F_{1}(x) = 0 \label{43}
\ee
determines the solution in a unique way. We find $F_{1}(x)= A_{1}(x)F_{0}(x)$,
where
\be
A_{1}(x) = \sqrt{\pi}(\tau - \tau^{-1})(x - \frac{1}{3}x^{3}) \label{44}
\ee
So, up to terms of order $\epsilon$, the distribution $F_{\epsilon}$ reads
\be
F_{\epsilon}(x)= \left[ 1 + \epsilon \sqrt{\pi}(\tau - \tau^{-1})
(x - \frac{1}{3}x^{3}) \right]F_{0}(x) \label{45}
\ee

The general frame developed here permits to continue the determination of 
higher order terms in a systematic way. The calculation of the second
order contribution $F_{2}(x)$ is presented in the appendix.

\section{Average velocity of the piston: concluding comments}

Using the definition (\ref{27}) we can evaluate the average velocity of
the piston in the stationary state associated with different temperatures
$T_{1},\; T_{2}$, but with the same pressures on both sides.
We find
\begin{eqnarray}
< V > & = & \sqrt{\frac{2k_{B}\sqrt{T_{1}T_{2}}}{M}}\int dx x \FE \label{46}\\
{} & = & \sqrt{\frac{2k_{B}\sqrt{T_{1}T_{2}}}{M}}\epsilon \sqrt{\pi}(\tau
- \tau^{-1})[ < x^{2} > - \frac{1}{3}< x^{4} >] \nonumber \\ 
{} & = & \frac{\sqrt{2\pi}}{4}\sqrt{\frac{m}{M}}\left[
\sqrt{\frac{k_{B}T_{2}}{M}} - \sqrt{\frac{k_{B}T_{1}}{M}} \right] \nonumber
\end{eqnarray}
Equation (\ref{46}) contains the fundamental prediction of our analysis.
It shows that the average velocity of the piston is different from zero
although the pressures are the same. Moreover, if $T_{2}>T_{1}$, the velocity
is positive showing that the piston evolves in the
direction of the high temperature region. In conclusion we thus recover in the more 
physical situation  of $M \gg m$, qualitatively the same result as
in the equal mass case $M=m$ studied in [5]. The advantage of the
present analysis is that the Boltzmann description is asymptotically
exact when $\epsilon =\sqrt{m/M}\to 0$, because the effect of recollisions
vanishes in this limit. We thus conclude that for a massive piston
a macroscopic motion is induced despite the equality of the pressures,
reflecting the asymmetry of fluctuations. This result, obtained for an
infinite volume, supports the point of view that the adiabatic piston
separating finite volumes of gases at equal pressures but different
temperatures will be also systematically displaced until the uniform
distribution is attained (for a rigorous analysis of the special case $M=m$ see [6])
The average kinetic energy of the piston is determined by
the zero order Maxwell distribution (\ref{40}),(\ref{400}), with
temperature $\sqrt{T_{1}T_{2}}$.
We thus have
\be
<\frac{M}{2}V^{2} > = \frac{k_{B}\sqrt{T_{1}T_{2}}}{2} \label{47}
\ee
It follows from (\ref{46}) and (47), that
\be
< V > = \sqrt{\frac{\pi m}{8M}}\left[ \left(\frac{T_{2}}{T_{1}}\right)^{1/4}-
\left(\frac{T_{1}}{T_{2}}\right)^{1/4}\right]\sqrt{< V^{2} >} \label{48}
\ee
Let us end this section by writing down the asymptotic formula for the
stationary distribution $\PH (V)$. When $\epsilon =\sqrt{m/M} \ll 1$,
\begin{eqnarray}
\PH (V) & = & \left\{ 1+\sqrt{\frac{m\pi}{2k_{B}}}\left[\frac{1}{\sqrt{T_{1}}}
-\frac{1}{\sqrt{T_{2}}}\right]\left(V-\frac{mV^{3}}{6k_{B}\sqrt{T_{1}T_{2}}}
\right)\right\} \label{49}\\
{} & \times & \sqrt{\frac{M}{2\pi k_{B}\sqrt{T_{1}T_{2}}}}
{\rm exp}\left(-\frac{MV^{2}}{2k_{B}\sqrt{T_{1}T_{2}}}\right) \nonumber
\end{eqnarray}\\[0.5cm]
%\newpage
\begin{center}
{\bf APPENDIX}
\end{center}
\underline{Order $\epsilon^{2}$}\\

From the expansion (\ref{36}) there follows an equation relating
$F_{2}(x)$ to $F_{0}(x)$ and $F_{1}(x)$ 
\be
[2J_{-1}x + 6J_{1}\frac{d}{dx}]F_{2}(x)+[J_{-2}x^{2}+16J_{2}\left(
\frac{d}{dx}\right)^{2}]F_{1}(x)+ \label{50}
\ee
\[ +[J_{-3}\frac{x^{3}}{3}+40J_{3}\left(\frac{d}{dx}\right)^{3}
+3J_{-1}x^{2}\frac{d}{dx}+16J_{1}x\left(\frac{d}{dx}\right)^{2}
+8J_{1}\frac{d}{dx}]F_{0}=0 \]
Putting $F_{2}(x)=A_{2}(x)F_{0}(x)$  we get with the help of (\ref{34})  a
first order equation
\be
\frac{d}{dx}A_{2}(x)=(\tau -\tau^{-1})^{2}[\pi (\frac{1}{3}x^{5}-
\frac{10}{3}x^{3}+4x)+(\frac{14}{3}x^{3}-8x)] \label{51}
\ee
The solution 
\be
A_{2}(x)= (\tau - \tau^{-1})^{2}[ {\cal A}+
(2\pi -4)x^{2}-\frac{1}{6}(5\pi -7)x^{4}+\frac{1}{18}\pi x^{6}] \label{52}
\ee
contains a constant ${\cal A}$ which can be determined from
the condition (\ref{380}). One finds
\be
{\cal A} = \frac{23}{48}\pi-\frac{9}{8} \label{53}
\ee
Hence, the second order contribution to $F_{\epsilon}(x)$ is given by
\be
F_{2}(x) = (\tau - \tau^{-1})^{2}[\frac{23}{48}\pi-\frac{9}{8}+
(2\pi -4)x^{2}-\frac{1}{6}(5\pi -7)x^{4}+\frac{1}{18}\pi x^{6}]
F_{0}(x)\label{54}
\ee
Let us remark that the general structure of the series expansion
(\ref{35}) has the product form of a polynomial in variable $x$
multiplying the gaussian distribution $F_{0}(x)$ (\ref{40}). At the order
$\epsilon^{3}$ one gets terms in $x^{5},x^{7},x^{9}$, at the
order $\epsilon^{4}$ terms in $x^{8},x^{10},x^{12}$ appear. In general,
the contribution of the order $\epsilon^{r}$ yields a polynomial
factor in $x^{3r-4},x^{3r-2}, x^{3r}$.

\section*{Acknowledgment}
J. Piasecki greatly acknowledges the hospitality at the Institute of Theoretical Physics of the \'Ecole Polytechnique F\'ed\'erale de Lausanne where this research has been performed.

\renewcommand{\baselinestretch}{1}

\end{document}